%% file: General.tex
\documentclass[10pt,conference,letterpaper]{IEEEtran}
\IEEEoverridecommandlockouts
\usepackage[utf8]{inputenc}
\usepackage{graphicx}
\usepackage{mathrsfs}
\usepackage{mathtools}  % amsmath with extensions
\usepackage{amsfonts}  % (otherwise \mathbb does nothing)
\usepackage{amsmath}
\usepackage{algorithm}
\usepackage{multicol}
\usepackage{textcomp}
\usepackage{algorithmicx}

\DeclareMathOperator*{\argmax}{arg\,max}

\usepackage{algpseudocode}
\usepackage[colorinlistoftodos]{todonotes}

\title{Online Frequency Scheduling by\\ Learning Parallel  Actions}
\author{Anastasios~Giovanidis$^{1*}$, Mathieu~Leconte$^{1*}$, Sabrine~Aroua$^1$, Tor~Kvernvik$^2$, David~Sandberg$^2$\\ $^1$Ericsson Research, Ericsson France, 25 Carnot, Massy 91300,
France\\ 
$^2$Ericsson Research, Ericsson AB, Torshamnsgatan 21, Kista,
Stockholm 164 83, Sweden\\
Email: \{firstname.lastname\}@ericsson.com}
\date{May 2024}

\begin{document}

\newcommand{\hif}[1]{{\color{blue}~{\bf [Anastasios:} #1]~}}

\maketitle
\def\thefootnote{*}\footnotetext{These authors contributed equally to the work.}
\input{Section/abstract}


%------ II.Intro --------
\section{Introduction}
\label{sec:intro}
\input{Section/introduction}

%------ II.Problem --------
\section{Problem Statement} \label{sec:problem}
\input{Section/Problem}

%------ III.Method -------
\section{Methodology} \label{sec:method}
\input{Section/model}

%------ IV.Unibranch -------
\section{Memory-efficient architectures} \label{sec:unibranch}
\input{Section/unibranch}

%------ V.Performance ------
\section{Performance Evaluation} \label{sec:perf}
\input{Section/Performance}

%------ VI.Conclusions -----
\section{Conclusion} \label{sec:conc}
\input{Section/Conclusion}

\bibliographystyle{IEEEtran}

\end{document}

%% file: Section/abstract.tex
\begin{abstract}
Radio Resource Management is a challenging topic in future 6G networks where novel applications create strong competition among the users for the available resources. In this work we consider the frequency scheduling problem in a multi-user MIMO system. Frequency resources need to be assigned to a set of users while allowing for concurrent transmissions in the same sub-band. Traditional methods are insufficient to cope with all the involved constraints and uncertainties, whereas reinforcement learning can directly learn near-optimal solutions for  such complex environments. However, the scheduling problem has an enormous action space accounting for all the combinations of users and sub-bands, so out-of-the-box algorithms cannot be used directly. In this work, we propose a scheduler based on action-branching over sub-bands, which is a deep Q-learning architecture with parallel decision capabilities. The sub-bands learn correlated but local decision policies and altogether they optimize a global reward. To improve the scaling of the architecture with the number of sub-bands, we propose variations (unibranch, GNN-based) that reduce the number of parameters to learn. The parallel decision making of the proposed architecture allows to meet short inference time requirements in real systems. Furthermore, the deep Q-learning approach permits online fine-tuning after deployment to bridge the sim-to-real gap. The proposed architectures are evaluated against relevant baselines from the literature showing competitive performance and possibilities of online adaptation to evolving environments. 
%Optimizing antenna tilts in cellular networks provides a practical and cost-efficient method to achieve a good tradeoff between network coverage and capacity.
%Previous methods based on Reinforcement Learning (RL) have shown promising results for tilt control by learning policies outperforming traditional tilt optimization solutions.
%However, most existing RL methods optimize the antenna tilt based on limited observation of the network state, which comprises the mean values of specific Key Performance Indicators (KPIs).
%In fact, the users’ distribution in the cell provides crucial information to decide the network coverage and control antenna tilt.
%In this paper, we propose RL-SFD, a novel RL-based Spatial Feature Detection framework, to control the antenna tilt. In RL-SFD, the antenna tilt is controlled based on how the users are distributed in the network in addition to their associated KPIs. We use Convolutional Neural Networks (CNN) to extract the spatial distribution of the users and obtain a richer latent representation of the network state. The RL network then exploits the output of the CNN to decide the antenna tilt that maximizes the reward function. Extensive simulation results show that RL-SFD efficiently captures relevant information and outperforms multiple baselines that don't include the users' distribution in the observation of the RL agent.

\end{abstract}

%% file: Section/introduction.tex
The road to 6G comes along with unique challenges raised by new types of applications, such as enhanced Mobile Broadband (eMBB) for high quality multimedia
services, interactive gaming, Augmented/Virtual Reality (AR/VR) and massive Ultra Reliable
Low Latency Communications (mURLLC). These enforce strong requirements for even higher data rates, lower latency and the support of a large and versatile number of mobile devices, rendering rule-based or optimisation-based methods inefficient and necessitating Artificial Intelligence (AI) empowered
solutions \cite{AIRAN6G24}.  Essentially, the telecom industry’s vision is
``AI-native'', i.e. the concept where AI is a natural part of
the network functionality, in terms of design, deployment,
operation, and maintenance \cite{ericsson_white_ainative}.

Radio Resource Management (RRM) plays an important role in satisfying the diverse future service requirements, by allocating dynamically time-frequency resources to User Equipments (UEs), with the aim to maximize some measure of performance (sum-throughput, fairness, etc). Further extensions of this challenging problem include solving for beamforming and power control, as well as link adaptation. To enhance the network's spectral efficiency, multiple UEs can be scheduled simultaneously over the same time-frequency blocks, by exploiting multiple antennas in both transmitter and receiver side, with the MU-MIMO paradigm \cite{RA-muMIMO17}. Another important challenge is that resource allocation decisions should be taken in real-time, within delays below $1\ [msec]$ for 5G and even lower for 6G \cite{OpenAirInterface_scheduling22}. The general RRM is known to be a mixed-integer problem \cite{ScheduleSurvey} and decisions need to be taken with uncertain state information. Considering also the potentially competing UE service requirements, the RRM task will become more and more complicated in the 6G and can be better handled by AI tools and more specifically by Reinforcement Learning (RL), which should also adapt to evolving environments (traffic mix, channel). The compute-heavy scheduling workloads from AI-based algorithms will require eventually the use of GPUs for hardware acceleration \cite{ericsson_silicon}, \cite{ericsson_intel}, \cite{Nvidia_ORAN}. A big challenge is then to adapt the size of AI models to available memory limitations, and to propose solutions whose inference time (e.g. forward pass of a neural network) requirements are a fraction of the slot duration. 

\subsection{Related work}
There exists a series of research works that apply RL to the RRM problem. Motivated by AlphaGo Zero the authors in \cite{DavidToricc22} propose a Monte Carlo Tree Search (MCTS) RL algorithm to solve the frequency scheduling with MU-MIMO user association. %The problem is combinatorial with a large number of solutions especially since it allows to schedule multiple users per frequency resource. 
The authors take sequential decisions over the frequency bands in a given order, and incorporate self-attention mechanisms to account for possibly interfering co-scheduled users. %Note that MCTS does not allow for online data collection from the real environment, because it needs to be trained on a simulator for the roll-out to run from the root to the leaves of the decision tree. 
In another work, LEASCH \cite{LEASCH} applies Double \textit{Deep Q-Network} (DQN) to learn the optimal time-frequency block allocation policy for a set of users using channel state, buffers, HARQ, and allocation logs as input. The algorithm simplifies considerably the decision process, by allowing at most one user per resource block (no MU-MIMO), and by deciding sequentially over blocks based on an input fairness metric, while assuming perfect knowledge over achievable rates. The work in \cite{RM6Grasti} applies a \textit{Dueling} DQN to solve the joint network selection and sub-band frequency allocation problem for both terrestrial and aerial UEs. The large number of users and carriers considerably increases the action-space and challenges the problem's feasibility. The authors in \cite{BoutibaSlice} solve the RRM time-frequency scheduling problem for slicing using DQN. The actions are the numerology choice and the number of resources per UE. This is a Layer-3 scheduling problem, without resource sharing among UEs and without explicit channel knowledge. DQN with \textit{Action-Branching} is applied in \cite{BranchPowerC} for uplink power control. The simultaneous optimization of two correlated variables is performed as two output parallel branch decisions. This approach is proposed to reduce the action-size of the problem instead of enumerating all possible action combinations from both variables. An interesting approach named GQN was introduced in \cite{BoutonGQN23} to remotely control the antenna-tilt and power for a number of base-stations in a cooperative manner. GQN combines graph neural networks (GNN) with per-agent DQN, so that the policy can learn from message-passing interactions between agents. The network can learn using a single global reward over all agents, whereas the global value function is decomposed among agents through value decomposition methods \cite{ValueDecomposition}.

\subsection{Challenges}
In this work we consider the Layer-2 problem of frequency scheduling in the downlink with MU-MIMO association, similar to \cite{DavidToricc22}. It is combinatorial with very large action space, which consists of all possible user allocations over all sub-bands. To give a numerical example, for just $4$ available users scheduled over $10$ sub-bands, with at most $2$ MU-MIMO users per sub-band, there are $10^{10}$ possible actions, rendering vanilla RL solutions intractable. The suggested approaches in \cite{DavidToricc22} and in \cite{LEASCH} (the latter without MU-MIMO) propose to solve the problem autoregressively over frequency bands, by associating users one sub-band at a time and then feeding the past decisions as input states for the next sub-band. The benefit of such approach is that it decomposes the general problem into smaller tractable sub-problems. At inference time (i.e. when the algorithm is deployed in a radio network) actions are produced per sub-band but without relying on MCTS which reduces the latency by several orders of magnitude. One drawback with the autoregressive approach is however that the inference time increases linearly with the number of sub-bands, since the learned network should be queried as many times as the number of sub-bands to yield a full scheduling decision. This is prohibitive however, because of the aforementioned tight delay requirements well below $1 [msec]$. 
Another drawback with the approach taken in \cite{DavidToricc22} relates to how training is performed. This approach relies on MCTS to compute policy targets to use for neural network training. This works very well in a non-realtime setting (e.g. training using interactions with a simulator) but due to the very high computational requirements from MCTS it is extremely challenging to apply in realtime settings. In many cases the UE receiver structure, and thereby the quality of a scheduling decision, is unknown at training time. It would therefore be beneficial to learn the rewards from interactions with real UEs.

\subsection{Our contributions}
We propose a Deep Q-Network (DQN) solution to the discrete scheduling problem that can overcome the above limitations by allowing for parallel (i.e. distributed) inference over all sub-bands simultaneously. We use the \textit{Action-Branching} (AB) network architecture \cite{Tavakoli_Pardo_Kormushev_2018}, where each sub-band is modelled as a separate branch that learns the locally optimal actions. Coordination over branches is achieved by learning a common shared representation of the joint input state, consisting of the complex channel state and buffer state of all UEs. An extra mechanism of coordination is introduced in the loss function by introducing value decomposition of the common reward among the individual branches \cite{ValueDecomposition}. Also, the relatively low compute requirements for training and the replay buffer architecture makes DQN applicable to cases where learning from realtime interactions is required.

The introduced approach shows high performance but uses large neural networks with a number of parameters that increases linearly with the number of branches. We investigate an alternative architecture where, after the shared representation block, a common sub-network (\textit{unibranch}) is used for both training and inference for all branches, thus reducing considerably the number of parameters and allowing for the architecture to scale for very large number of branches. The common network uses as extra input the preprocessed local state (i.e. user channels on specific sub-band and their buffers) as well as positional encoding for the branch label information. 

We make a final step by replacing the shared representation block with a Graph Neural Network (GNN) \cite{deepGNNs}, where we assume that a sub-band is a node on a graph whose edges are their pairwise interactions. The GNN takes as input the local state, like in the unibranch case above, and outputs a learned local representation for each branch. The latter is fed into a common DQN network for all branches, to learn the optimal local actions, combined together using value decomposition. %The approach is similar to the GQN \cite{BoutonGQN23}.

The remainder of the paper is organised as follows.
Section~\ref{sec:problem} formalises the scheduling problem under study.  Section~\ref{sec:method} introduces the RL solution based on action-branching, which allows for reduced inference latency due to parallelism. Then, Section~\ref{sec:unibranch} presents the memory efficient architectures with unibranch and GNNs, which use a smaller network but a part of the inference is done sequentially.  The experimental environment, the baselines used for comparison, as well as the performance evaluation are presented in Section~\ref{sec:perf}. All solutions are compared in terms of their performance and achieved trade-off between inference time and network size. The possibility for online adaptation of the trained DQN architectures to realtime environments is showcased.   Finally, Section~\ref{sec:conc} concludes our work. 

%% file: Section/Problem.tex
We consider the problem of scheduling a number $N_u>1$ of users over time-frequency blocks. We take decisions over discrete time slots $t=1,\ldots$ each with duration $\Delta_t\ [msec]$. There are $N_s>1$ available sub-bands of size $\Delta_f\ [Hz]$ with carrier frequency $f_0\ [Hz]$ and total bandwidth $BW = N_s\Delta_f\ [Hz]$. This is a sequential decision making problem over time, but the association over all sub-bands should be done simultaneously per slot. We describe it as a Markov Decision Process (MDP) characterised by the tuple $\{\mathcal{S}, \mathcal{A}, \mathcal{P}, \mathcal{R}, \gamma\}$.

\textit{States:} $\mathcal{S}$ is the set of all joint UE states per time slot. Each user provides information about their instantaneous estimated channel, as well as their buffer state. The channel state for all users is a complex tensor with dimensions $(N_u\times N_s \times N_{rx} \times N_{tx})$, where $N_{rx}$ is the number of receive UE antennas and $N_{tx}$ the number of station transmit antennas. The estimation diverges from the exact channel due to noise in measurement, as well as UE mobility. The buffer state is an $N_u$-dimensional vector with positive integers from $0$ to a maximum buffer size.

\textit{Actions:} The set of actions $\mathcal{A}$ is discrete and refers to all possible combinations of associating the $N_u$ users to the $N_s$ sub-bands. A user may be associated in general to more than one sub-band. The number of possible user associations per sub-band, given at most $M\geq 1$ co-channel transmissions in MU-MIMO, is equal to $N_a= \sum_{x=1}^{N_u}\binom{N_u}{x}$. E.g. for $N_u=4$ users and $M=2$ there are $N_a=10$ possible associations. 
The global action set over all sub-bands is  $\mathcal{A}=\mathcal{A}^{(1)}\times\ldots\times \mathcal{A}^{({N_s})}$  with cardinality $|\mathcal{A}|=N_a^{N_s}$. As a numerical example, for  $N_u=4$ users that need to be scheduled over $N_s=10$ sub-bands and $M=2$ the action set has $|\mathcal{A}|= 10^{10}$ possible actions. In realistic scenarios we may need to decide over $N_u=10$ users over $N_s=20$ sub-bands.

\textit{Reward:} The reward function $\mathcal{R}:\mathcal{S}\times \mathcal{A}\rightarrow \mathbb{R}_+$ maps the current state-action pair $(s,a)$ to one-dimensional real positive reward $r$, related to some scheduling performance metric over all sub-bands and all users. In our work the reward per time slot is the sum of Throughput-To-Average over all users \cite{ScheduleSurvey}, that quantifies myopic Proportional Fair (PF) decisions. The reward is calculated as in \cite{DavidToricc22}. First a precoder is chosen per user and sub-band based on the estimated channel per slot. Next, the resulting Signal to Interference plus Noise Ratio (SINR) per user and sub-band is derived and their Transport Block Size (TBS) is computed by a simple link adaptation algorithm. Finally, the achieved rate per user $R_k$ is calculated by scaling the TBS with the success probability (one-minus Block Error Probability). The PF utility function used is 
\begin{eqnarray}
\label{PF}
r & = & \nu \sum_{k=1}^{N_u}\frac{R_k}{\overline{R}_k}.
\end{eqnarray}
In the above $R_k$ and $\overline{R}_k$ are the instantaneous and average rate for user $k$, where the average rate uses the channel slow-fading values and a random allocation. $\nu$ is a normalising constant. The above metric quantifies the relative rate advantage of allocating specific sub-bands to a user compared to an average allocation and channel realisation. It is very relevant for Mobile Broadband (MBB) traffic whose quality of experience (QoE) depends on the bitrate. For delay sensitive services utility metrics based on delay would be more appropriate.

\textit{Transitions:} The function $\mathcal{P}:\mathcal{S}\times \mathcal{A}\times\mathcal{S}\rightarrow [0,1]$ describes the transition probability to the next state $s'$, given the current state-action pair $(s,a)$ and is unknown in real environments. This transition quantifies channel correlation as well as buffer evolution due to service and random arrivals.

\textit{Discount:} The discount factor $\gamma\in[0,1]$ weights the importance of future rewards in the cumulative sum of rewards over a long horizon.

We are looking for an optimal stationary user association policy
$\pi:\mathcal{S}\times\mathcal{A}\rightarrow [0,1]$ to maximise the expected cumulative discounted reward from any initial state $s_0\in\mathcal{S}$ at time $t=0$. The \textit{value function} and \textit{state-action value function} are
\begin{eqnarray}
\label{Svalue}
V_{\pi}(s_{0}) & := & \mathbb{E}_{\pi}\left[\sum_{t=0}^{\infty}\gamma^tr_{t}|s_{0}\right],\\
\label{SAvalue}
 Q_{\pi}(s_{0},a_{0}) & := & \mathbb{E}_{\pi}\left[\sum_{t=0}^{\infty}\gamma^tr_{t}|s_{0},a_{0}\right].
\end{eqnarray}

The reason why we chose the specific type of myopic reward in eq.(\ref{PF}) is because the frequency scheduling problem is in practical cellular systems distinguished from time scheduling. The latter is a separate function which keeps track of the bitrate and delay experienced by each user and schedules them aiming for specific performance intents over time. The same reward was also suggested in \cite{DavidToricc22} for these reasons. To apply the MDP to such myopic problem we can choose $\gamma=0$ for the discount factor and the problem simplifies to finding the optimal solution to a very large discrete optimisation problem over given channel and buffer states. However, we need to underline that the MDP formulation (and the DQN solution approach in the next sections) is more versatile. %In practice the frequency scheduler will apply its decisions within a window of time slots. %, before the time scheduler chooses the next set of users to be served. 
The MDP can learn the long-term optimal frequency allocation policy within a window of time-slots, by setting the $\gamma$ factor appropriately, this being a major advantage compared to myopic solutions.

%% file: Section/model.tex
\subsection{Deep Q-Network}
In real wireless environments the transition probabilities are unknown, whereas the reward function cannot be explicitly given due to very large state space and non-ideal user channel estimation. Instead, by probing the state the corresponding reward can be sampled. Note that the reward can be fed back with some delay in online settings, which is not critical for training as we will clarify later. To solve the MDP for a real environment of channel conditions and request arrivals, we can use a model-free approach, where the optimal actions can be learned using value-based methods such as Q-learning \cite{RL}. Such choice of method is appropriate for discrete action sets. 

To account for very large state spaces (complex channels and buffers) a parameterised value function $Q(s,a;\mathbf{\theta})$ should be learned, where $\mathbf{\theta}$ are parameters of a deep network that inputs the current system state $s$ and outputs the vector over all possible state-action values. This deep network is tuned by experience and can generalise to unseen states. The Deep Q-Network approach \cite{mnih2013human} optimally tunes the $\mathbf{\theta}$ parameters by minimising over the expected temporal difference $TD(0)$-error
\begin{eqnarray}
\label{DQNloss}
L(\mathbf{\theta}) & = & \mathbb{E}_{s,a\sim \mathcal{\rho}}\left[\left(y-Q(s,a;\mathbf{\theta})\right)^2\right],\\
\label{target}
y & = & \mathbb{E}_{r, s'\sim \mathcal{E}}\left[r+\gamma \max_{a'\in{\mathcal{A}}}Q^-(s',a';\mathbf{\theta^-})|s,a\right].
\end{eqnarray}
The $(s,a)$ pairs are sampled from a behavior distribution which trades-off between exploration and exploitation. The standard option is the $\epsilon$-greedy policy, where with probability $\epsilon(t)$ a random action is selected, while with probability $1-\epsilon(t)$ the greedy action is applied. The exploration probability cools-down as the time $t$ evolves. The distribution $\mathcal{E}$ is over all possible rewards $r$ and next states $s'$ the environment can randomly transition to from $(s,a)$, depending on the environment dynamics. It should be noted that due to the relatively low computational complexity of the DQN these updates can be executed in a base station in real time. 

\textbf{Prioritized Experience Replay:} %In practice, the behaviour policy and transitions are sampled from $\rho$ and $\mathcal{E}$ respectively. 
At step $t$ the samples from $\rho$ and $\mathcal{E}$ form an experience $e_t=(s_t,a_t,r_t,s'_{t+1})$ that is pushed in a list of fixed length $L$, the so-called replay memory $\mathcal{D}$ \cite{mnih2013human}. At each loss-gradient update of the parameters, a batch of $B$ experiences is sampled at random from the memory, thus avoiding strong correlations between samples, and allowing each experience to be re-used in many updates. We make use of prioritized experience replay \cite{PrioritizedReplay}, where samples with high learning progress are more frequently
sampled from the memory. Priority per experience $(s_t,a_t,r_t,s'_{t+1})$ is quantified by the absolute $TD(0)$ difference. %$|r(s_t,a_t)+\gamma\max_{a'\in{\mathcal{A}}}Q(s_{t+1},a';\mathbf{\theta})-Q(s_{t},a_t;\mathbf{\theta})|$ 
Sampling from memory is controlled by two exponents $(\alpha,\beta)$ related to prioritization and importance sampling, respectively.

\textbf{Online Updates:} In an online setting the result from a downlink transmission after associating a user to a sub-band will be available to the scheduler once the UE has acknowledged a reception and the result has been fed back to the base station over the Hybrid Automatic Repeat reQuest (HARQ) feedback channel. Hence, an extra delay is inherent in the way the reward is calculated, when taking Block Error Probability (BLEP) into account as in our case. This delayed outcome can be handled by updating the replay buffer in two steps: $(s_t,a_t)$ are updated first as soon as the action is chosen and $(r_t,s'_{t+1})$ are updated in a second step when the reward and next state are fed back. Experience replay and random sample updates permit such flexibility. 

\textbf{Target network:} In eq.(\ref{target}) the state-action value to find the maximum and its  parameters are marked by a minus ($-$). This denotes a target network, whose parameters are updated by soft iterations  slower than the actual Q-network. This approach can offer stabilization in DQN convergence \cite{mnih2013human}.

\subsection{Action Branching}
The big challenge in the multi-user frequency scheduling problem is the enormous number of actions, equal to $|\mathcal{A}|=N_a^{N_s}$, as explained in Section \ref{sec:problem}. Such numbers of actions render the classical DQN method infeasible. 

The approach we take here is inspired by the multi-agent literature \cite{tampuuMultiAgent17}. Each sub-band $d=1,\ldots,N_s$ can be considered an agent with action space $\mathcal{A}^{(d)}$ of cardinality $|\mathcal{A}^{(d)}|=N_a$ and should decide over the best local action, given some input. In that case, the total number of actions for the problem can be reduced to $N_sN_a$, e.g. for the toy example, the total actions now reduce to just $10\cdot 10=100$ (i.e. $10$ per sub-band). The big challenge, however, is that the agents should cooperate in order to maximize their joint reward $\mathcal{R}$ given in the MDP of the original problem. An important extra benefit of this approach is the \textit{parallelisation} of the decision making among the agents, who can all decide simultaneously about their local action, thus reducing the delay in the forward pass, compared to sequential implementations over agents.%, such as the MCTS approach in \cite{DavidToricc22}.

A good candidate method for discrete actions is the \textit{action branching}, introduced in \cite{Tavakoli_Pardo_Kormushev_2018}. The main idea of the proposed architecture is to distribute the action-making across individual network branches (one agent aka branch per sub-band), but at the same time, maintain a common shared latent representation of the input to help with the coordination of the branches.

\begin{figure}[t!]
    \centering
\includegraphics[width=8.5cm]{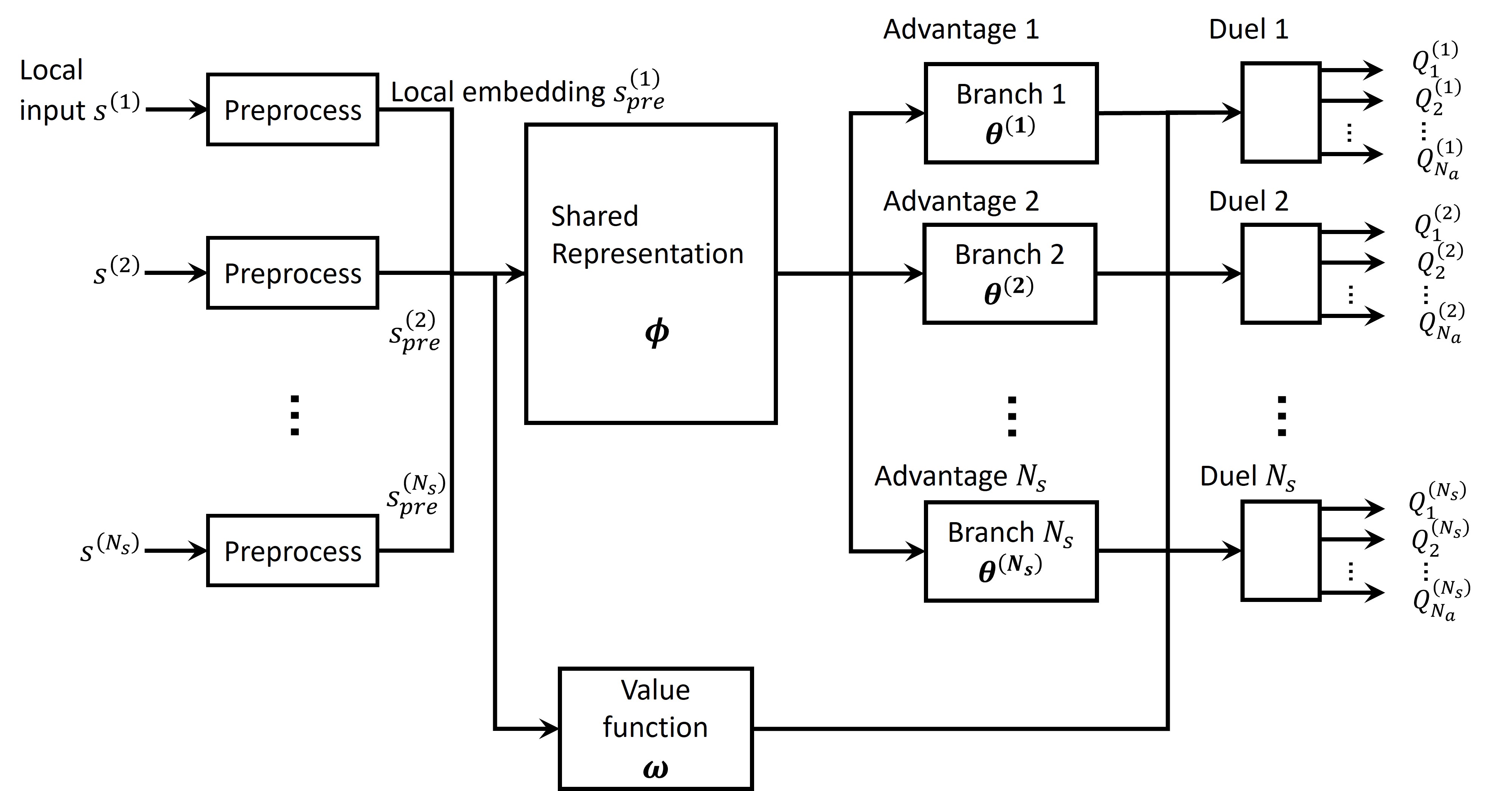}
    \caption{Action branching architecture}
    \label{fig:AB architecture}
\end{figure}

The proposed network architecture is shown in Fig.\ref{fig:AB architecture}. It consists of the following blocks:
%\begin{enumerate}
%\item[i.] 
(i.) Preprocessing, 
%\item[ii.] 
(ii.) Shared representation, 
%\item[iii.] 
(iii.) Value-function, 
%\item[iv.] 
(iv.) Advantage block per sub-band, 
%\item[v.] 
(v.) Dueling block per sub-band.
%\end{enumerate}

\subsection{Detailed Architecture}

The input consists of a batch of $B$ joint states of complex channels for all $N_u$ users over the $N_s$ sub-bands, and their buffer state. The complex channel input has dimension $(B,2N_u,N_s,N_{rx},N_{tx})$, where the real and imaginary parts are concatenated, $N_{rx}$ is the number of receive antennas at the user side and $N_{tx}$ the number of transmit antennas at the base station.

\textbf{Preprocessing:} The input is preprocessed separately for each sub-band as introduced in \cite{DavidToricc22}. The channel information is transformed into engineered features aiming to quantify how much co-scheduled users could interfere with each other. They represent the channel of each user-pair by three scalars, the magnitude of the dot product between the channel vectors of user pairs, the Hermitian angle and Kasner’s pseudo angle. The output of this preprocessing step is a flat 1-dimensional vector of size $N_u N_{feat}$ for each sub-band, with number of features per user and sub-band $N_{feat}=2+3N_u$, including the normalized buffer size and channel power. These vectors then go through a dense layer to obtain a fixed-sized embedding for the input of each sub-band,
\begin{eqnarray}
    \label{preproc}
    s_{pre}^{(d)}=p(s^{(d)}), & & d=1,\ldots,N_s.
\end{eqnarray}

\textbf{Shared Representation:} The feature vectors of each sub-band are then concatenated as $s_{pre}$ and enter the shared representation module (ii), which is a multi-layer perceptron (MLP) with linear layers and ReLU activation. The set of its parameters is $\mathbf{\phi}$. The output is a common embedded state of size $N_{hidden}$ used as input for all branches. 
\begin{eqnarray}
    \label{embed}
    s_{emb}(\phi) = f_{\text{MLP}}(s_{pre};\phi). 
\end{eqnarray}

\textbf{Branching with Dueling:} We apply the dueling method \cite{dueling} with action branching, to produce a separate estimate of the value function $V(s)$ and a separate estimate for the advantage functions. This architecture has
been shown to lead to better policy evaluation in the presence
of actions with similar state-action values. The value function module (iii) is an MLP block with parameters $\omega$, input $s_{pre}$ and scalar output. For each of the $d=1,\ldots,N_s$ parallel branches the common embedded state $s_{emb}$ is input to each local MLP block (iv.) with parameters $\theta^{(d)}$ that learns the advantage function $A^{(d)}(s_{emb},a^{(d)})$ for all possible local actions $a^{(d)}\in\mathcal{A}^{(d)}$. Following the proposal in \cite{dueling}, we obtain the local state-action value $Q^{(d)}$ by aggregating the advantage and value functions at the (v.) dueling block as follows
\begin{eqnarray}
\label{advantageQ}
& & Q^{(d)}(s,a^{(d)};\phi,\omega,\theta^{(d)})  =  A^{(d)}(s,a^{(d)};\phi,\theta^{(d)}) + \nonumber\\
& & \mspace{20mu} + \left(V(s;\omega)-\frac{1}{|\mathcal{A}^{(d)}|}\sum_{a'\in\mathcal{A}^{(d)}}A^{(d)}(s,a';\phi,\theta^{(d)})\right).
\end{eqnarray}
In the above, subtracting the average of advantages instead of the maximum increases the stability of the optimization.
%We denote the ensemble of network parameters by 
%\begin{eqnarray}
%\label{params}
%$\Theta = \left\{\phi,\omega,\left\{\theta^{(d)}\right\}_{d=1}^{N_s}\right\}$.
%\end{eqnarray}

\subsection{Loss function}

As mentioned, we will focus on the myopic scheduling case, where the discount factor is $\gamma=0$, in order to compare with other myopic frequency-schedulers in the literature. The temporal difference target in (\ref{target}) now simplifies to $y=r$, and we need not discuss ways to combine the $N_s$ branches of $Q^{(d)}$ in the reward-to-go. In \cite{Tavakoli_Pardo_Kormushev_2018} the loss was defined as the expected value of the mean squared TD error averaged over all branches $d$, i.e. the expected value of the quantity $\frac{1}{N_s}\sum_{d=1}^{N_s}\left(r-Q^{(d)}(s,a^{(d)})\right)^2$. %,
%\begin{eqnarray}
%    \label{ll_loss}
%    L_{//}(\Theta) = \mathbb{E}_{(s,a,r,*)\sim\mathcal{D}}\left[\frac{1}{N_s}\sum_{d=1}^{N_s}\left(r-Q^{(d)}(s,a^{(d)})\right)^2\right].
%\end{eqnarray}
 %We index the loss above by $//$ to denote that 
This way each branch approximates the joint total reward $r$ with its local $Q^{(d)}$ value.  We have found empirically that this approach under performs.

\textbf{Value Decomposition (VD):} It is more beneficial to assume that each of the local $Q^{(d)}$ values contributes partially to the total reward $r$. Hence, the idea of VD for cooperative multi-agent learning \cite{ValueDecomposition} is applied, which states that the global action value function in eq.(\ref{SAvalue}) can be additively decomposed into local value functions among the branches $Q(s,a)\approx \sum_{d=1}^{N_s}Q^{(d)}(s,a^{(d)})$. The loss now takes the form
\begin{eqnarray}
    \label{VDN}
    L_{VD}(\Theta) = \mathbb{E}_{(s,a,r,*)\sim\mathcal{D}}\left[\left(r-\sum_{d=1}^{N_s}Q^{(d)}(s,a^{(d)})\right)^2\right].
\end{eqnarray}
where the experience $(s,a,r,*)$ is sampled from the prioritized replay buffer $\mathcal{D}$, but the next state $s'$ need not be considered in the myopic case (so we mark by $*$). The $a=(a^{(1)},\ldots,a^{({N_s})})$ is the concatenation of local actions over all sub-bands. 

In practice the factorization in eq.(\ref{VDN}) ensures that a set of parallelized individual $\arg\max$
operations performed over  individual $Q^{(d)}$-functions to select greedily the optimal action per branch, yields the same result as a global $\arg\max$ on the global Q-function over the joint agent
actions $a=(a_1,\ldots,a_{N_s})$. Notice that there are many powerful alternatives to the VD, such us the QMIX \cite{QMIX} which factorizes the global Q-value assuming monotonicity, and the Deep Coordination Graphs \cite{DeepCoordGraphs}, \cite{bouton2021coordinated}, which factorize not only into individual utility Q-functions but additionally using pairwise payoff functions among agents. Here, we found VD to combine simplicity with performance.

%% file: Section/unibranch.tex
The action branching architecture may be difficult to scale to large number of parallel decisions, because it requires a lot of parameters. The two main reasons are that the number of parameters of the branches $N_s\cdot \theta^{(d)}$ scales linearly with the number of branches, and that the shared representation has to contain useful information for all the branches at the same time and will thus tend to be large. In this section, we evolve from the basic action branching architecture to attain better-scaling, more parameter-efficient architectures. The first evolution is to have the branches share parameters, which we call the unibranch architecture. The second evolution will avoid the shared representation and let each decision dimension build its own local representation by employing a Graph Neural Network (GNN) architecture.

\subsection{Unibranch}

\begin{figure}[t!]
    \centering
\includegraphics[width=8.5cm]{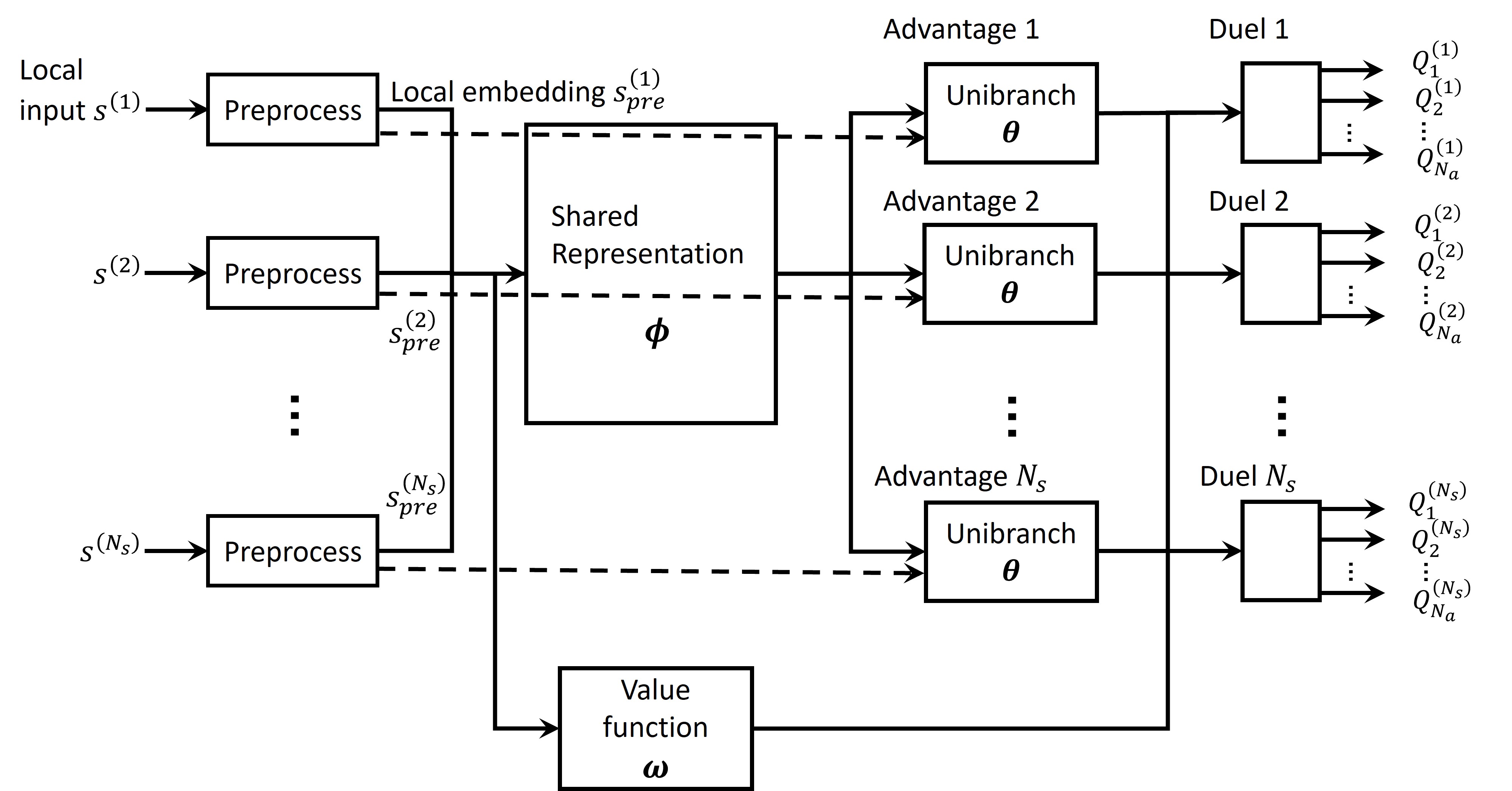}
    \caption{Unibranch architecture}
    \label{fig:Unibranch architecture}
\end{figure}

The advantage function in the action branching architecture was $A^{(d)}(s,a^{(d)};\phi,\theta^{(d)})=A^{(d)}(s_{emb}(\phi),a^{(d)};\theta^{(d)})$. Sharing the parameters of the branches by simply having $\theta^{(d)}=\theta, \forall d,$ would lead to all the branches taking the same decision. Therefore, we add a second input to each branch, that is branch-dependent. The preprocessing step was actually computing a separate embedding for each branch input: $s_{pre}^{(d)}=g(s^{(d)})$, $d=1,\ldots,N_s$. We concatenate this branch-specific state $s_{pre}^{(d)}$ to the shared representation $s_{emb}(\phi)$ to form the branch input:
\begin{eqnarray}
    \label{unibranch}
    A^{(d)}(s,\cdot;\phi,\theta)=f_{\text{MLP}}([s_{emb}(\phi), s_{pre}^{(d)}];\theta), & & \forall d.
\end{eqnarray}
The unibranch is again implemented using an MLP block. Its input is slightly larger than in the initial action branching architecture, as the local branch-specific input $s_{pre}^{(d)}$ was added, yet the shared representation is typically much larger than this local input, so the gain obtained by sharing parameters across branches still largely dominates. The unibranch architecture is shown on Figure~\ref{fig:Unibranch architecture}.

\subsection{Graph Neural Networks}

\begin{figure}[t!]
    \centering
\includegraphics[width=9cm]{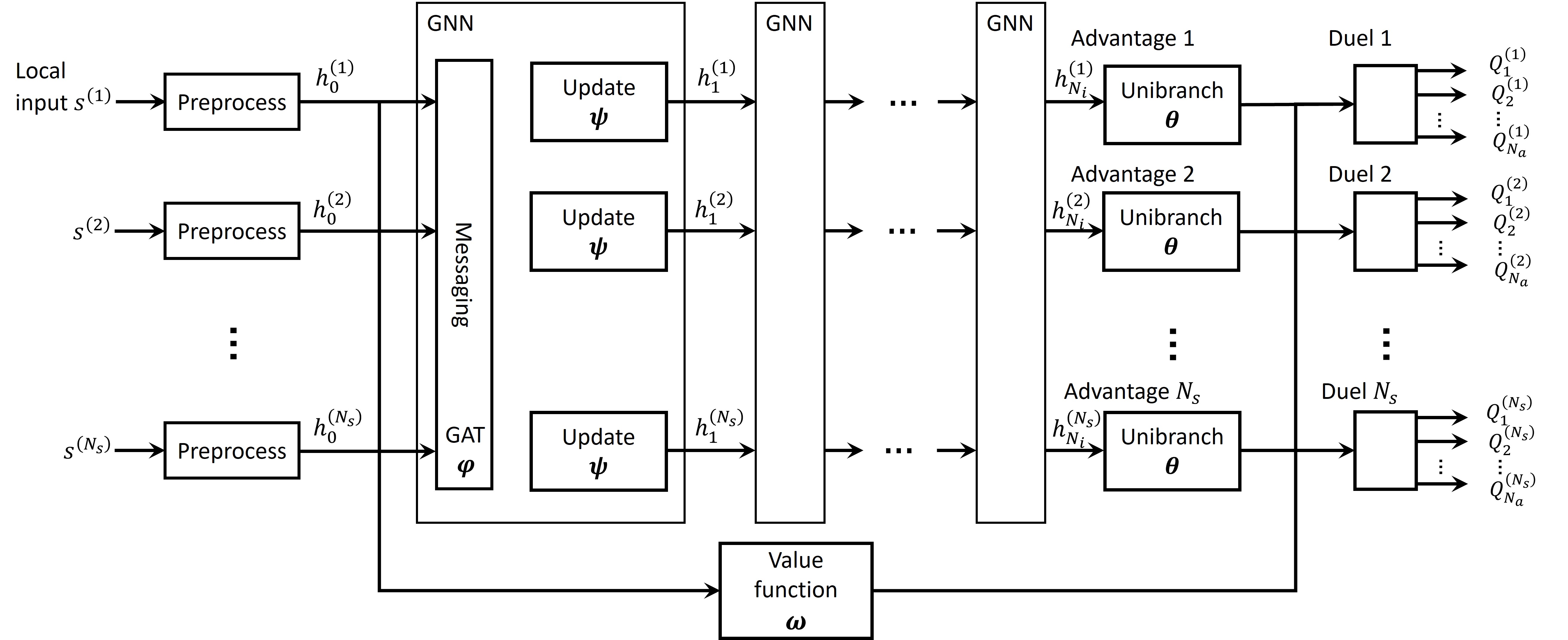}
    \caption{GNN architecture}
    \label{fig:GNN architecture}
\end{figure}

The action branching and unibranch architectures both suffer from the relatively large size of the shared representation. Indeed, it needs to capture all the information relevant to all the decision dimensions. In order to reduce further the size of the architecture, we move to only using local representations for each action dimension. However, these local representations still need to enable coordination of the decisions. To this end, we use a Graph Neural Network (GNN) architecture, which allows the local representations to be jointly updated over multiple iterations $N_i$. For the GNN architecture, we used a Graph Attention Network (GAT) \cite{velivckovic2017graph} layer followed by an MLP layer to update the representation. The parameters $(\varphi, \psi)=\phi$ of these layers play the same role as the parameters of the shared representation update of the previous architectures, yet the local representations $h_i^{(d)}$ involved in this GNN architecture are much smaller than the shared representation. The advantage function is then computed as
\begin{eqnarray}
    \label{GNN}
    A^{(d)}(s,a^{(d)};\phi,\theta)&=&A(h^{(d)}_{N_i},a^{(d)};\theta)\nonumber\\
    h^{(d)}_{i+1}&=&f_{\text{GNN}}(h_i^{(d)},h_i^{(-d)};\phi)\nonumber\\
    &=&f_{\text{MLP}}\left(f_{\text{GAT}}(h_i^{(d)},h_i^{(-d)};\varphi);\psi\right)\nonumber
\end{eqnarray}
Having more GNN iterations leads to more coordinated local representations, yet increases inference time, hence a trade-off needs to be found. The GNN architecture is shown in Figure~\ref{fig:GNN architecture}.

%% file: Section/Performance.tex
%\begin{itemize}
% \item List of DQN parameters for both Action-branching and GNN
% \item List of simulation parameters
%\item evaluate training convergence rate (to emphasize difference especially from the GNN approach which should be slower)
% \item memory-size model comparison
%\item performance comparison among models related to PF metric
%\item forward response time/delay of the models by evaluation
%\item Possible experiment: train for a number of specific users and evaluate for a different number of users, to see how much each of the models can generalise. Similar with user velocity and channel uncertainty.
%\end{itemize}

\subsection{Experimental setup}\textbf{Network Simulator}: We use a similar performance evaluation setup as in \cite{DavidToricc22}. The setup represents an urban macro-cell as a circle sector of radius 500 m and central angle 65°. Users are dropped at random in this area, at least 35 m away from the antenna. Their channel model is described by 3GPP TR 38.901 \cite{38901}. The buffers of the users are drawn uniformly at random as $N_{bits}\sim U(b_{min},b_{max})$. For more details on the experimental setup, the reader can refer to \cite{DavidToricc22}. The environment parameters are shown in Table \ref{tab:envrionment parameters}. We use the same number of users and sub-bands as the evaluation in  \cite{DavidToricc22}.

    \begin{table}[h]
    \caption{Environment parameters}
    \centering
    \begin{tabular}{l|c|c}
    Parameter & Symbol & value or size \\\hline\hline
    Nr. users & $N_u$ & $4$\\
    Nr. frequency bands & $N_s$ & $10$\\
    Nr. actions per sub-band & $N_a$ & 10\\
    Maximum nr. users per sub-band & $M$ & 2\\
    Channel Model & & 3GPP Urban Macro\\
    Carrier Frequency & & 3.5 GHz\\
    Deployment & & Single Cell\\
    Transmit antennas & $N_{tx}$ & 2\\
    Receive antennas & $N_{rx}$ & 1\\
    Slot duration & $T_{slot}$ & 1 ms\\
    Minimum buffer (bits) & $b_{min}$ & 400 \\ 
    Maximum buffer (bits) & $b_{max}$ & $\in\{\infty, 4000, 2000, 1000\}$\\
    User speed (m/s) & $v$ & $\in\{0, 1, 3, 5\}$\\
    Channel estimation quality (dB) & $\text{SNR}_{CE}$ & $\in\{\infty, 20, 10, 0\}$
    \end{tabular}
    \label{tab:envrionment parameters}
    \end{table}

%The simulation is executed on an urban network consisting of $B$ base stations, $C$ cells and $U$ UEs randomly positioned in the environment. The simulation parameters are reported in Table I.

\textbf{Evaluated architectures}: The parameters of the evaluated architectures are shown in Table~\ref{tab:architecture parameters}. For all the MLPs involved, the size of the hidden layers evolves linearly between input and output. This leads to the architecture sizes in Table~\ref{tab:architecture size}. 

    % \begin{table}[h]
    % \caption{Architecture design parameters}
    % \centering
    % \begin{tabular}{l|c|c}
    % Parameter & Symbol & value or size  \\\hline\hline
    % Local input & $s^{(d)}$ & 56\\
    % Local representation & $h_i^{(d)}, s_{pre}^{(d)}$ & 128 \\
    % Shared representation & $s_{emb}$ & 1280\\
    % Shared representation: nr. MLP layers & & 3\\
    % Branches/Unibranch: nr. MLP layers & & 2\\
    % GNN messaging: nr. GAT heads & & 3\\
    % GNN representation update: nr. MLP layers & & 2\\
    % GNN iterations & $N_i$ & 3\\
    % \end{tabular}
    % \label{tab:architecture parameters}
    % \end{table}
    
    \begin{table}[h]
    \caption{Architecture design parameters}
    \centering
    \begin{tabular}{l|c|c}
    Parameter & Symbol & value or size  \\\hline\hline
    Local input & $s^{(d)}$ & 56\\
    Local representation & $h_i^{(d)}, s_{pre}^{(d)}$ & 64 \\
    Shared representation & $s_{emb}$ & 640\\
    Shared representation: nr. MLP layers & & 3\\
    Branches/Unibranch: nr. MLP layers & & 2\\
    GNN messaging: nr. GAT heads & & 3\\
    GNN representation update: nr. MLP layers & & 2\\
    GNN iterations & $N_i$ & 3\\
    \end{tabular}
    \label{tab:architecture parameters}
    \end{table}

    % \begin{table}[h]
    % \caption{Size of the architectures}
    % \centering
    % \begin{tabular}{l|c}
    % Architecture & Number of trainable parameters \\\hline\hline
    % Action Branching & $14\cdot 10^6$\\
    % Unibranch & $7\cdot 10^6$ \\
    % GNN & $10^5$\\
    % \end{tabular}
    % \label{tab:architecture size}
    % \end{table}

    \begin{table}[h]
    \caption{Size of the architectures vs Inference time}
    \centering
    \begin{tabular}{l|c|c}
    Architecture & trainable parameters & relative inference time \\\hline\hline
    alphaZero \cite{DavidToricc22} & $5.2\cdot 10^4$ & $16.3$ \\
    Action Branching & $3.6\cdot 10^6$ & $1$ \\
    Unibranch & $1.7\cdot 10^6$ & $1.07$ \\
    GNN & $2.8\cdot 10^4$ & $1.76$ \\
    \end{tabular}
    \label{tab:architecture size}
    \end{table}

\textbf{Training process}: To benefit from parallelization of the DQN architecture in sample generation and decision evaluation, we train the policy in multiple iterations. In each iteration, we generate 1000 new environment samples and draw actions according to an $\varepsilon$-greedy exploration policy, with a decaying value of $\varepsilon$. After evaluating the reward for these samples, they are added to a replay buffer of size $L=10^5$. At each iteration, we run 100 optimization steps with batches of 256 samples from the replay buffer. Additional training parameters are mentioned in Table~\ref{tab:training parameters}.

% \begin{table}[h]
% \begin{center}
%   \caption{The Neural Network Parameters}
%   \begin{tabular}{ l  c  c c}
%     \hline
%      Layer & Shared  & Advantage & Value  \\\hline\hline
%     \hline
%      Input channels & $XX$ & $XX$ & $XX$\\
%     \hline
%     Output channels & $XX$ & $1$ & $XX$\\ \hline
%   \end{tabular}
%    \end{center}
% \end{table}

    % \begin{table}[h]
    % \caption{Training parameters}
    % \centering
    % \begin{tabular}{l|c|c}
    % Parameter & Symbol & value or size  \\\hline\hline
    % Total number of training samples & & $10^5$\\
    % Total number of optimization batches & & $2.5\cdot 10^4$\\
    % Batch size & B & 256\\
    % Experience replay memory size & $L$ & $10^4$\\
    % Discount factor & $\gamma$ & $0$\\
    % Experience replay prioritization & $\alpha$ & 0.7 \\
    % Experience replay sampling exponent & $\beta$ & 0.5\\
    % Optimizer & & AdamW\\
    % Learning rate & & $10^{-4}$\\
    % Exploration method & & $\varepsilon$-greedy\\
    % \end{tabular}
    % \label{tab:training parameters}
    % \end{table}

    \begin{table}[h]
    \caption{Training parameters}
    \centering
    \begin{tabular}{l|c|c}
    Parameter & Symbol & value or size  \\\hline\hline
    Total number of training samples & & $10^6$\\
    Total number of optimization batches & & $10^5$\\
    Batch size & B & 256\\
    Experience replay memory size & $L$ & $10^5$\\
    Discount factor & $\gamma$ & $0$\\
    Experience replay prioritization & $\alpha$ & 0.7 \\
    Experience replay sampling exponent & $\beta$ & 0.5\\
    Optimizer & & AdamW\\
    Learning rate & & $10^{-4}$\\
    Exploration method & & $\varepsilon$-greedy\\
    \end{tabular}
    \label{tab:training parameters}
    \end{table}

%--------------
\subsection{Baseline policies} In our experiments we compare the proposed solution against two baselines, one traditional and one alphaZero-based, which are both described below.

\textbf{Traditional Baseline: }\label{baseline_scheduler}
The traditional baseline scheduler used in this work is based on the optimization method defined by \cite[Alg. 7.1]{ResourceAllocBook}, which is a strong heuristic algorithm with quadratic complexity. Some modifications are introduced to support scheduling of up to $M$ users per sub-band (see Alg.~\ref{baseline_scheduler_alg}).

Sub-bands are allocated to users in a way that maximizes the marginal utility ($\Lambda_{k,j}$), i.e. the gain in the utility $U_{PFTF,k}$ when an extra sub-band $j$ is allocated to user $k$, compared to the utility of user $k$ before the allocation of sub-band $j$. The optimization algorithm (including modifications) is outlined in Algorithm~\ref{baseline_scheduler_alg}. Scheduling decisions are represented by $\mathcal{I}_{s,k}$ corresponding to the set of sub-bands allocated to user $k$ and $\mathcal{I}_{u,j}$ corresponding to the set of users allocated to sub-band $j$. The transmit power is split equally between sub-bands and between the users allocated to a sub-band. The maximum number of co-scheduled users $M$ is set to $2$ in this work.

The instantaneous rate for user $k$ is calculated by scaling the transport block size (TBS) with the success probability and dividing by the slot duration $T_\text{slot}$ as 
\begin{equation}\label{throughput_eq}
R_k = \frac{(1-\text{BLEP}_k)*\min(\text{TBS}_k,N_{bits,k})}{T_\text{slot}}.
\end{equation}
Here, the block error probability $\text{BLEP}_k$ is again calculated using the method in~\cite{l2s}, $N_{bits,k}$ is the number of bits in the buffer for user $k$ and $T_\text{slot}$ is the slot duration.% which is 1 ms in our scenario.

As optimization criterion we use the Proportional Fair Time Frequency (PFTF) metric %~\cite{pftf} 
which can be written as
\begin{equation*}%\label{pftf_metric_eq}
    U_{PFTF,k} = 
\begin{cases}
    \frac{\sum\limits_{j\in\mathcal{I}_{s,k} \cup \{i\} }{R_{k,j}^{}}}{\overline{R}_k+\sum\limits_{j\in\mathcal{I}_{s,k}}{R_{k,j}^{}}},& \sum\limits_{j\in\mathcal{I}_{s,k}}{R_{k,j}T_{slot}^{}} \leq N_{bits,k} \\
    0,              & \text{otherwise}
\end{cases}
\end{equation*}
where $R_{k,j}$ is the rate for user $k$ in sub-band $j$ in the current slot based on Eq.~\eqref{throughput_eq}, $\overline{R}_k$ is the average rate of user $k$ over a time window excluding the current slot. It should also be noted that the utility for user $k$ is set to $0$ when the buffer for the user is emptied, i.e. when the number of bits for user~$k$ exceeds its buffer size $N_{bits,k}$.

\begin{algorithm}[t!] \scriptsize
\caption{Traditional Baseline Algorithm}\label{baseline_scheduler_alg}
\begin{algorithmic}[t!]
\State $\mathcal{I}_{avail} \in \{j \:|\: j \in 1 \dots N_{s}\}$
\State $\mathcal{I}_{s,k} \gets \{\}$ $\forall k \in 1 \dots N_{u}$
\State $\mathcal{I}_{u,j} \gets \{\}$ $\forall j \in 1 \dots N_{s}$
\While{$\mathcal{I}_{avail} \neq \{\}$}
\For{$j \in \{i \in 1\dots N_{s}|\:|\mathcal{I}_{u,i}| < M\}$}
\For{$k \in 1 \dots N_{u}$}
\State \textbf{Compute} $U_{PFTF,k}(\mathcal{I}_{s,k} \cup \{j\})$
\If{$\mathcal{I}_{s,k}\neq \{\}$}
\State \textbf{Compute} $U_{PFTF,k}(\mathcal{I}_{s,k})$
\Else
\State $U_{PFTF,k}(\mathcal{I}_{s,k}) \gets 0$
\EndIf
\State $\Lambda_{k,j} \gets U_{PFTF,k}(\mathcal{I}_{s,k} \cup \{j\}) - U_{PFTF,k}(\mathcal{I}_{s,k})$
\EndFor
\EndFor
\State $(k^*, j^*) \gets \argmax_{(k,j)} \Lambda_{k,j}$
\If{$\Lambda_{k^*,j^*} > 0$}
\State $\mathcal{I}_{s,k^*} \gets \mathcal{I}_{s,k^*} \cup \{j^*\}$
\State $\mathcal{I}_{u,j^*} \gets \mathcal{I}_{u,j^*} \cup \{k^*\}$
\Else
\State $\mathcal{I}_{avail} \gets \mathcal{I}_{avail} \setminus \{j^*\}$
\EndIf
\EndWhile
\end{algorithmic}
\end{algorithm}

\textbf{alphaZero-based Baseline: }\label{mcts_scheduler}
We also compare to the alphaZero-based approach outlined in \cite{DavidToricc22}. In this approach, a neural network is trained using policy and value targets from Monte Carlo Tree Search. Training is done by alternating between two phases. In a first phase, MCTS with 1000 simulations is used to find actions to maximize the associated rewards. In a second phase, a neural network is trained to predict the action probabilities and the corresponding values. These two phases are iterated for 25 iterations, with 500 episodes in each iteration.

%--------------
\subsection{Results \& Discussion}

We view the scenario with perfect channel estimation ($\text{SNR}_{CE}=\infty$), infinite buffers ($b_{max}=\infty$), and no user mobility ($v=0$) as a reference. We let only one parameter at a time deviate from this reference scenario. Table~\ref{tab:PF perf} shows the performance in terms of PF metric compared to the baseline scheduler. The performance reported in these tables is based on $10^4$ samples. Recall that the baseline and alphaZero schedulers take sequential decisions, while the Action Branching, Unibranch, and GNN schedulers are taking parallel decisions.

    % \begin{table}[h]
    % \caption{Relative performance over traditional baseline with limited buffers}
    % \centering
    % \begin{tabular}{|l||c|c|c|c|}
    % \hline
    % $b_{max}$ & alphaZero \cite{DavidToricc22} & Action Branching & Unibranch & GNN \\\hline\hline
    % $\infty$ & 106\% & 106\% & *\% & 107\% \\\hline
    % 4000 bits & 103\% & 105\% & *\% & *\% \\\hline
    % 2000 bits & 97\% & 100\% & *\% & *\% \\\hline
    % 1000 bits & 98\% & 100\% & *\% & 99\% \\\hline
    % \end{tabular}
    % \label{tab:PF perf buffer}
    % \end{table}

    \begin{table}[h]
    \caption{Relative performance over traditional baseline with various (top) buffers, (middle) velocity, (bottom) SNR estimation}
    \centering
    \begin{tabular}{|l||c|c|c|c|}
    \hline
    $b_{max}$ & alphaZero \cite{DavidToricc22} & Action Branching & Unibranch & GNN \\\hline\hline
    $\infty$ & 102\% & 104\% & 107\% & 105\% \\\hline
    4000 bits & 112\% & 110\% & 112\% & 110\% \\\hline
    2000 bits & 118\% & 119\% & 120\% & 115\% \\\hline
    1000 bits & 131\% & 130\% & 135\% & 126\% \\\hline
    % &  & & &  
    %\end{tabular}
    %\label{tab:PF perf buffer}
    %\end{table}
    %
    % \begin{table}[h]
    % \caption{Relative performance over traditional baseline with user velocity}
    % \centering
    % \begin{tabular}{|l||c|c|c|c|}
    % \hline
    % $v$ & alphaZero \cite{DavidToricc22} & Action Branching & Unibranch & GNN \\\hline\hline
    % 0 & 106\% & 106\% & *\% & 107\% \\\hline
    % 1 m/s & 108\% & 106\% & *\% & *\% \\\hline
    % 3 m/s & 168\% & 164\% & *\% & *\% \\\hline
    % 5 m/s & 174\% & 172\% & *\% & 170\% \\\hline
    % \end{tabular}
    % \label{tab:PF perf speed}
    % \end{table}
    %
    %\begin{table}[h]
    %\caption{Relative performance over traditional baseline with user velocity}
    %\centering
    %\begin{tabular}{|l||c|c|c|c|}
    \hline\hline
    $v$ & alphaZero \cite{DavidToricc22} & Action Branching & Unibranch & GNN \\\hline \hline
    0 & 102\% & 104\% & 107\% & 105\% \\\hline
    1 m/s & 117\% & 97\% & 120\% & 123\% \\\hline
    3 m/s & 189\% & 155\% & 179\% & 182\% \\\hline
    5 m/s & 194\% & 160\% & 184\% & 187\% \\\hline
    % & & & & 
    %\end{tabular}
    %\label{tab:PF perf speed}
    %\end{table}
    %
    %\begin{table}[h]
    %\caption{Relative performance over traditional baseline with imperfect channel estimation}
    %\centering
    %\begin{tabular}{|l||c|c|c|c|}
    \hline\hline
    $\text{SNR}_{CE}$ & alphaZero \cite{DavidToricc22} & Action Branching & Unibranch & GNN \\\hline \hline
    $\infty$ & 102\% & 104\% & 107\% & 105\% \\\hline
    20 dB & 100\% & 102\% & 107\% & 106\% \\\hline
    10 dB & 133\% & 127\% & 133\% & 133\% \\\hline
    0 dB & 262\% & 244\% & 256\% & 257\% \\\hline
    \end{tabular}
    \label{tab:PF perf}
    \end{table}

\begin{figure*}[t!]
    \centering
\includegraphics[width=5.5cm]{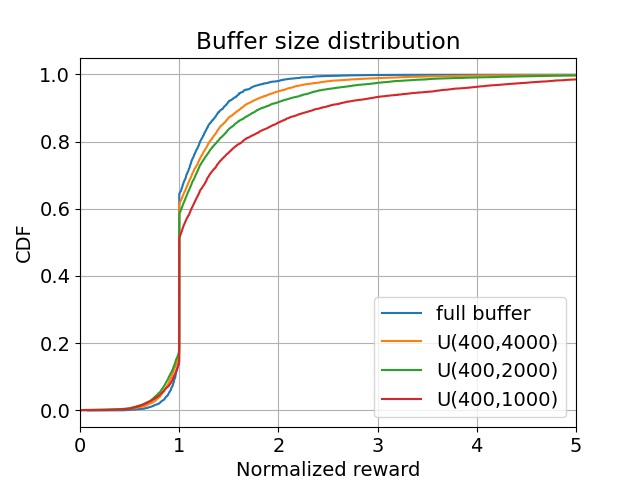}
\includegraphics[width=5.5cm]{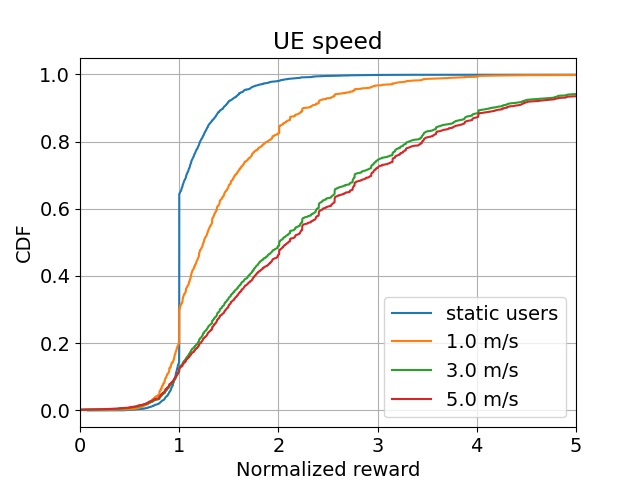}
\includegraphics[width=5.5cm]{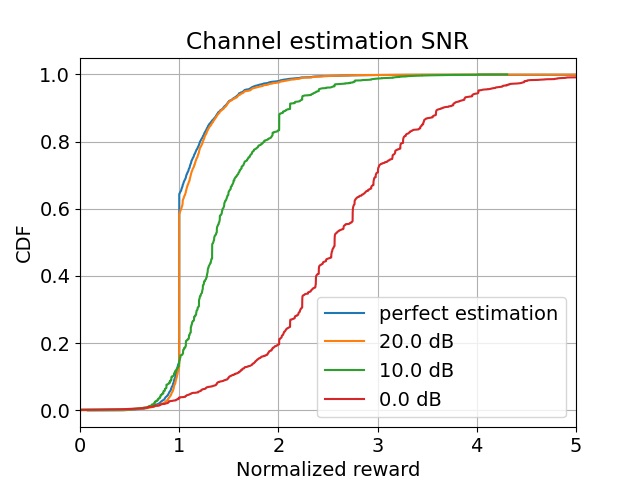}
    \caption{CDFs of the rewards obtained by the GNN architecture divided by the respective baseline reward, obtained over $10^4$ validation episodes. Environment variations over (left) buffer size, (middle) user speed, (right) SNR estimation. The other DQN-based parallel architecture variations show similar curves.}
    \label{fig:cdf}
\end{figure*}

Table~\ref{tab:PF perf} (Top) shows the performance with limited buffers. In these scenarios there is no uncertainty on the state, but the traditional baseline scheduler does not properly coordinate the amount of spectral resources allocated to a user with the quantity of data in that user's buffer. All the evaluated methods exceed baseline. The benefits become more visible as the buffer sizes decrease, which requires more coordination among sub-bands.
Table~\ref{tab:PF perf} (Medium) shows the performance with user velocity. At 1 m/s speed, the channel aging slightly affects the predicted channels, while at 3 m/s and 5 m/s the predicted channels widely differ from the true ones. All the machine learning methods provide substantial gain over the traditional baseline in these scenarios. The Action Branching seems a bit behind compared to Unibranch and GNN.
Table~\ref{tab:PF perf} (Bottom) shows the performance with imperfect channel estimation. The results show the same trend as with user velocity, with higher benefits for lower SNR values.

Overall, the proposed parallel decision architectures manage to take scheduling decisions of competitive quality with the alphaZero-based approach, largely improving over the baseline. %The Action Branching architecture shows worse performance than the other ML-based schedulers.
The comparison of Action Branching with Unibranch suggests that training one common branch may be more efficient than training a different sub-network per branch. Though much smaller in number of parameters, the GNN scheduler manages to reach the same level of performance as the other two parallel architectures, but at the price of more message-passing iterations ($N_i>1$), and thus a larger inference time while still smaller than the sequential decisions of the alphaZero-based approach, as we will see next.

In Fig.~\ref{fig:cdf} we plot the cumulative distribution function (CDF) of the rewards from the GNN architecture divided by the baseline. The plots show that the reward improvement over baseline spreads over a large range depending on the scenario realization, with only a few cases at the left of the vertical line $x=1$ where the baseline is better. 
%TODO
%Compare the policies in terms of PF performance, memory size, inference time (delay) and training time. 

%Discuss the variations of the result with channel noise and user speed.

%We can discuss generalisation capabilities for environments different than the one where the algorithm is tested (number of users, channel noise, speed etc)

\textbf{Network Size vs Inference time trade-off:} Comparing the size of the models from Table~\ref{tab:architecture size}, we can see that the parameter sharing in the branches allows the Unibranch architecture to halve the number of parameters compared to Action Branching (AB) without performance loss as shown above, yet the largest gain in number of parameters is obtained by the GNN architecture, which avoids the large shared representation.
In Table~\ref{tab:architecture size}, we also show the inference times we obtain for the different architectures. The inference times are evaluated from 300 runs on the same GPU. We normalize the inference times by that of the Action Branching architecture, which is the lowest. We observe that, although the alphaZero architecture seems competitive in size with the GNN, it suffers however from very high inference time, $16$-times larger than the AB. The Unibranch %with half the parameters of the AB 
has almost the same inference time as AB. The GNN exhibits very interesting properties by managing in just $1.7$ times the delay of AB to provide the same performance with two orders of magnitude fewer parameters.  %The objective here is to show the relative differences between the architecture to highlight the trade-offs between them.

    % \begin{table}[h]
    % \caption{Relative performance over traditional baseline with imperfect channel estimation}
    % \centering
    % \begin{tabular}{|l||c|c|c|c|}
    % \hline
    % $SNR_{CE}$ & alphaZero \cite{DavidToricc22} & Action Branching & Unibranch & GNN \\\hline\hline
    % $\infty$ & 106\% & 106\% & *\% & 107\% \\\hline
    % 20 dB & 106\% & 104\% & *\% & *\% \\\hline
    % 10 dB & 115\% & 117\% & *\% & *\% \\\hline
    % 0 dB & 288\% & 279\% & *\% & 266\% \\\hline
    % \end{tabular}
    % \label{tab:PF perf channel estimation}
    % \end{table}

\begin{figure}[t!]
    \centering
\includegraphics[width=8.5cm]{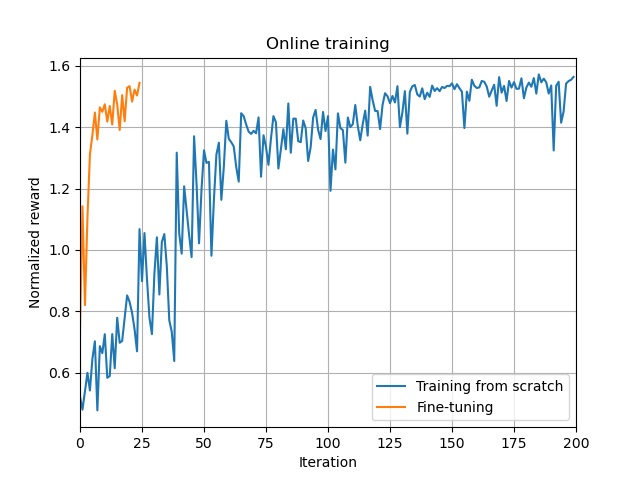}
    \caption{Training performance of fine-tuning vs. training from scratch, evaluated on 500 validation episodes.}
    \label{fig:online training}
\end{figure}

\textbf{Online Adaptation:} To test the ability of the proposed methods to quickly adapt after deployment, we train the GNN-version of our scheduler for a specific scenario: 10 users (56 actions per sub-band) moving at 1 m/s. We then "deploy" it on a related but different scenario: users now move at 3 m/s. The model is fine-tuned on the deployment scenario using new samples. Fig.~\ref{fig:online training} shows the performance evolution during this fine-tuning compared to a new training from scratch. We can see that our DQN parallel decision-making scheduler manages to quickly fill the initial performance loss caused by the scenario mismatch. Note that such fine-tuning would not be possible for the alphaZero scheduler of \cite{DavidToricc22}. For a TTI of 1 ms, then for i.i.d. data our algorithm adapts in 25 s to the new environment, compared to 3 minutes if trained from scratch. 

%% file: Section/Conclusion.tex
In this work we have introduced new reinforcement learning architectures that can efficiently learn the optimal user association for MU-MIMO frequency scheduling using parallel decision making. The first main benefit compared to sequential baselines %the sequential architectures of alphaZero-based solutions as well as greedy heuristics 
is the inference speed that allows to take scheduling decisions within strict delay requirements, without any loss in performance. Another important benefit is that DQN-based algorithms do not depend on a simulator to roll-out the policy but can rather explore-exploit online from collected data, thus learning in realtime and adapting the policy to changes in the environment. We proposed variations of the parallel architecture %where part of the learned network is common among the branches (unibranch, GNN) 
and showcased the trade-off between network size in number of variables and inference time, for the same performance in the target metric. %Another very important benefit of the DQN architectures compared to the alphaZero-based solution is the ability  to learn online in realtime by exploring actions directly on the UEs that are present, without the need of training on a simulator. In this way the proposed DQN networks can adapt to evolving environments. 

The proposed architectures have been evaluated for the myopic case with instantaneous reward. %where the reward is an instantaneous throughput metric and no delay or fairness over time is taken into account, hence the next state transition is not important. 
However, for traffic mixes where the algorithm should schedule users with different Quality-of-Experience (QoE) requirements (delay, throughput, reliability) the optimal scheduling policy should be learned jointly over longer time horizons, in which case the DQN algorithms are expected to show advantages compared to myopic solutions. Such temporal aspects studied within multi-cell scenarios are very challenging topics for future investigations. %This needs to be studied in future work by comparison against alphaZero or other myopic policies whose decisions are guided over time by some temporal scheduling policy. Another important  extension can consider interference-aware multi-cell scheduling. 